# Atomically-thin Femtojoule Filamentary Memristor

Huan Zhao[1], Zhipeng Dong[2], He Tian[1], Don DiMarzio[3], Myung-Geun Han[4], Lihua Zhang[5], Xiaodong Yan[1], Fanxin Liu[6], Lang Shen[1], Shu-jen Han[7], Steve Cronin[1], Wei Wu[1], Jesse Tice[3], Jing Guo[2,*], Han Wang[1,*]

[1]Ming Hsieh Department of Electrical Engineering, University of Southern California, Los Angeles, CA 90089, USA

[2]Department of Electrical and Computer Engineering, University of Florida, Gainesville, FL 32611, USA.

[3]NG Next, Northrop Grumman Aerospace Systems, 1 Space Park, Redondo Beach, CA 90278, USA.

[4]Condensed Matter Physics and Materials Sciences Department, Brookhaven National Laboratory, Upton, NY 11973, USA

[5]Center for Functional Nanomaterials, Brookhaven National Laboratory, Upton, NY 11973, USA

[6]Center for Optics & Optoelectronics Research (COOR), Collaborative Innovation Center for Information Technology in Biological and Medical Physics, and College of Science, Zhejiang University of Technology, Hangzhou 310023, PR China.

[7]IBM T. J. Watson Research Center, Yorktown Heights, NY 10598, USA

*To whom correspondence should be addressed: han.wang.4@usc.edu (H.W.), guoj@ufl.edu (J.G.)

The morphology and dimension of the conductive filament formed in a memristive device are strongly influenced by the thickness of its switching medium layer. Aggressive scaling of this active layer thickness is critical towards reducing the operating current, voltage and energy consumption in filamentary type memristors. Previously, the thickness of this filament layer has been limited to above a few nanometers due to processing constraints, making it challenging to further suppress the on-state current and the switching voltage. Here, we study the formation of conductive filaments in a material medium with sub-nanometer thickness, formed through the oxidation of atomically-thin two-dimensional boron nitride. The resulting memristive device exhibits sub-nanometer filamentary switching with sub-pA operation current and femtojoule per bit energy consumption. Furthermore, by confining the filament to the atomic scale, we observe current switching characteristics that are distinct from that in thicker medium due to the profoundly different atomic kinetics. The filament morphology in such an aggressively scaled memristive device is also theoretically explored. These ultra-low energy devices are promising for realizing femtojoule and sub-femtojoule electronic computation, which can be attractive for applications in a wide range of electronics systems that desire ultra-low power operation.

Resistive switching devices such as oxide based memristors[1-7] and conductive bridge random access memories (CBRAM)[8-10] utilize filamentary switching to realize low power memory and computation operations. In these devices, the formation and destruction of the conductive filament enables the reversible switching of the device conductivity,[11-13] resulting in compact two-terminal devices well suited for miniaturization and integration.[14-15] These devices are promising candidates for applications in low-current, low power memory and neuromorphic electronics.[16-23] However, the operating current in resistive switching devices are constrained by the scaling of the medium layer thickness that dictates the dimensions and morphology of the resulting filament.[2, 24-27] The thickness of the oxide based medium layer in existing low current memristive devices, typically deposited by atomic layer deposition method (ALD), is limited to above 3 nm[28-30] and the devices typically operate at the nano-ampere current level with the lowest record being 100 pA[31]. New materials and device designs are needed to form ultrathin medium layers that is critical for reducing the filament length to the atomic scale, and for minimizing the energy required to form and rupture them.

The recent emergence of two-dimensional materials has offered a new pathway for achieving a high quality atomically-thin medium layer to realize ultra-low power filamentary switching. In this work, we demonstrate the formation of a memristive switching medium layer with sub-nanometer thickness created through the oxidation of the atomically-thin two-dimensional (2D) hexagonal boron nitride. The reduction of the medium layer thickness allows the demonstration of sub-nanometer filamentary switching with sub-pA operation current and femtojoule per bit energy consumption. Moreover, fundamentally new atomic kinetics can arise when the medium layer thickness reduces to the atomic scale. The unique switching characteristics in this ultra-thin medium filamentary switching device is explained through Kinetic Monte Carlo simulations of the filament morphology and the distinct electrostatics in the atomically-thin medium.

To create the atomically-thin switching medium, few layer hexagonal boron nitride (h-BN) flakes were first exfoliated onto the Si/SiO$_2$ substrate with 285 nm SiO$_2$ via a standard mechanical exfoliation method. The layer thicknesses were initially estimated by optical contrast and further confirmed by Atomic Force Microscopy (AFM). We conducted Raman spectroscopy on all the few-layer h-BN samples. A characteristic Raman peak near 1367 cm$^{-1}$ corresponding to the E$_{2g}$ phonon vibration mode is observed in all the samples (Figure 1a). The few-layer samples were subsequently oxidized via an ultra-low power oxygen plasma treatment, which can oxidize the h-BN flakes up to a few nanometers in depth without damaging the flake morphology (see Methods). AFM topographic images were obtained to verify that the flakes are atomically smooth after oxygen plasma treatment and have approximately the same thickness as the sample before the treatment. Raman measurements showed that the h-BN Raman peak completely vanished after the oxygen plasma treatment, which suggests the full amorphization of h-BN. Atomic resolution scanning transmission electron microscopy (STEM) reveals that pristine h-BN thin films have crystallized "honeycomb" lattice structure and hexagonal electron diffraction patterns, while oxidized h-BN flakes are amorphous (Figure 1b). Furthermore, electron energy loss spectroscopy (EELS) and elemental quantification identified that the oxidized few-layer hBN sample contains boron, nitrogen, and oxygen, among which the oxygen element dominates (Figure 1c). Therefore we use $BNO_x$ to denote the amorphized boron oxide material in the following text.

Each $BNO_x$ flake was deterministically transferred from the substrate onto a few-layer graphene flake which serves as a thin, smooth and conformal bottom electrode (BE) of the memristor structure. 40 nm silver was subsequently deposited as the top electrode (TE), capped with another 40 nm gold to facilitate electrical probing (Figure 1a). Multiple devices were fabricated with $BNO_x$ layer thickness ranging from 0.9 nm to 2.3 nm, corresponding to bilayer to 5-layer h-BN before oxidation. The left panel of Figure 1d shows the AFM image of a 0.9 nm thick $BNO_x$ flake stacked on a 5-layer graphene BE and the corresponding AFM height profile. The right panel is a cross-section STEM image of the memristive device fabricated with this $BNO_x$ sample. We can clearly see a ~0.9 nm amorphous $BNO_x$ layer sandwiched between the graphene bottom electrode and the Ag top electrode.

Electrical measurements were conducted on devices with 0.9 nm, 1.3 nm, 1.8 nm, and 2.3 nm $BNO_x$ medium layer thickness under the current compliances of 5 pA, 90 pA, 500 pA, and 2 nA, respectively. The current compliance we applied for each device is the lowest current that can robustly form and rupture the filament while avoiding filament overgrowth. All the devices exhibit typical bipolar resistive switch behaviors in their I-V characteristics (Figure 2a). By applying a positive/negative voltage from BE to TE, a conductive filament of Ag atoms can be formed/ruptured, resulting in the switching to the low/high resistance state of the device. Once the positive voltage applied is larger than a critical voltage, i.e. the SET voltage, the Ag cations electrically connect the BE and TE, resulting in an abrupt jump of the current and therefore turning on the device. The migration of the Ag cations occurs at a relatively lower bias and dominates the conductance between the TE and BE. After a device is set, it will remain at the low resistance state (LRS) until a negative voltage is applied to reset the device to a high resistance state (HRS). The on/off ratio of our devices ranges from 100 to 1000, and this asymmetrical on-state current, which had enlarged the memory window and reduced the current level for rupturing the filament, is a signature of conductive bridge memristors.[8, 31] All these devices can reliably operate for at least hundreds of set/reset cycles in the ambient environment. The electrical characteristics of these memristive devices with atomically-thin switching layer demonstrated two distinctive features compared to the conventional thicker memristors. In previously reported memristive devices, the reset process is believed to form a depletion gap in the filament that electrically disconnects the TE and BE, while the set process re-connects the gap and therefore turns the device on. Several experiments have measured the filament depletion gap length to be ~6 nm[32-34], while a lower limit of 2-3 nm gap size was predicted via theoretical analysis[35]. In conventional memristive devices, the filament, which is longer than the gap distance, is generated through a forming step requiring a significantly larger voltage than that used for setting the device. In contrast, the thickness of all our switching layers are smaller than the size of conventional filament gaps, and therefore no forming step is needed and the set voltage scales proportionally with the $BNO_x$ switching layer thickness.

Another interesting feature of the $BNO_x$ devices is the strong dependence of its switching characteristics on the thickness of the $BNO_x$ layer. Devices with a thinner $BNO_x$ layer can be reliably operated under lower current and voltage in both the "set" and "reset" states. In particular, the set voltage increases linearly with the thickness of the $BNO_x$ layer (Figure 2b). Furthermore, when the voltage increases at the positive side, the current for the 0.9 and 1.3 nm samples show no notable increase until the abrupt jump occurs, while the current for the 1.8 nm and 2.3 nm samples increases gradually as the voltage increases until the voltage reaches the threshold to set the device, which is similar to conventional memristive devices.

To understand the unique thickness dependent properties of the switching characteristics, we carried out kinetic Monte Carlo (KMC) simulations. The KMC simulations describes cationic generation, hopping and reduction behaviors at a microscopic level, which provides qualitative understanding of the electrical characteristics in these aggressively scaled memristive devices. Figure 3a shows the filament morphology for the different switching medium thicknesses, $t_{ox}$, in which the blue dots denote the energy minimum sites in the $BNO_x$ layer; the yellow dots denote the active electrode atoms or cations and the gray dots denote the inert electrode atoms. It shows that an extremely thin filament down to a single atomistic chain can be formed between the electrodes for device with $t_{OX} \approx 0.9$ nm oxidized from bilayer hBN. In contrast, in the case of $t_{OX} \approx 4.5$ nm, because the filament grows in both the lateral and vertical directions before a percolative filament path can be formed, the filament expands much wider than a single atomistic chain in its lateral dimensions. Hence, the percolative filament path formed through a thicker oxide in the SET process would typically have a wider diameter.

In addition, an enhanced positive feedback process facilitates the thin filament formation in the device for the case of $t_{OX} \approx 0.9$ nm. For a resistive switching device with BNO$_x$ thickness $t_{OX}$ and an applied voltage of $V_0$, the electric field increases from $\frac{V_0}{t_{OX}}$ to $\frac{V_0}{t_{OX}-a_0}$ if a cation (yellow dot) reaches the bottom electrode through hopping as shown in Figure 3b. Here $a_0 \approx$ 0.45 nm is the grid spacing, which is treated as an approximate value for layer spacing in layered materials. For $t_{OX} \approx 0.9$ nm and $V_0 \approx 630$ mV, the vertical electric field approximately increases from ~0.7 V/nm to ~1.4 V/nm after the first cation hops into the switching medium, which results in an exponential increase in the local hopping rate and hence facilitates the subsequent vertical filament formation. In comparison, for $t_{OX} \approx 2.3$ nm and $V_0 \approx 1.6$V, the vertical electric field approximately increases from ~0.7 V/nm to ~0.86 V/nm after a cation hops into the switching medium during the initial filament formation stage. The change is much less significant compared to the vertical electrical field enhancement in the 0.9 nm switching layer case. The enhancement of the vertical growth rate, which has an exponential dependence on the vertical electric field, thereby, is much smaller. The larger increase in the vertical electric field in devices with ultrathin BNO$_x$ layer leads to more pronounced positive feedback process of vertical filament growth, which enhances the formation of a narrow atomistic filament chain. Furthermore, through KMC simulation, we found that the intrinsic set time reduces as the $BNO_x$ layer thickness decreases (see Figure S1 in the supplementary information) since it is easier to form a percolative path in a thinner $BNO_X$ layer. In addition, the number of hopping cations required to electrically connect the BE and TE becomes considerably smaller when the switching medium thickness scales down since not only the filament becomes shorter, but the lateral dimensions of the filament also becomes smaller (Figure 3a). Hence, the atomically-thin memristive device is expected to have a faster intrinsic operation speed than the memristors with thicker switching medium.

Figure 4 shows the characteristics of the thinnest memristive device we have demonstrated with 0.9 nm $BNO_x$ layer. Figure 4a shows the set and reset I-V characteristics of the device with 0.9 pA, 5 pA, and 9 pA current compliances. The high resistance state (HRS) current level fluctuates around 10-100 fA with no obvious gradual increase until the abrupt rise in the current occurs. The 10 -100 fA current level is dominated by the equipment noise. The device can be written and erased at sub-pA current level with a on/off current ratio ~10. The energy consumption for the programming (SET) and erasing (RESET) operations are estimated to be less than 10 fJ and 1 fJ, respectively, when we applied 10 ms voltage pulses to execute these operations. The device is expected to operate well with voltage pulses much shorter than 10 ms, which is the speed limit of our existing high sensitivity current measurement system. The set voltages under different current compliance are reliably near 0.6-0.7 V, which is consistent with the aforementioned Ag cation migration mechanism. Figure 4b shows the data retention of the memory device with a 0.9 nm $BNO_x$ switching layer, measured at 85°C. The device resistance at the low resistance state (LRS) was measured after the device was switched on under 5 pA current compliance, and HRS resistance was measured after the device was switched off. The data was read at 0.18 V without applying current compliance, with a total measurement time of 4 hours. The device shows excellent data retention, suggesting that the Ag filament is robust even for the operation at low power and elevated temperature. Figure 4c shows HRS and LRS current levels obtained during 100 continuous set-reset DC cycles, with a reading voltage of 0.15 V and current compliance of 5 pA. The set voltages statistics for the 100 cycles exhibit a normal distribution centered at 0.63 V and a standard deviation of 0.088 V (Figure 4d), indicating that the filament formation process is highly reproducible and consistent, even operated at ultra-low power. The device has a record low operation current that is about two orders of magnitude lower than previously reported[8]. The device is attractive for application as an ultralow power memristive computational device, for example as a binary synapse for neuromorphic computing.

In summary, the resistive switching of conductive filaments in a sub-nanometer thickness material medium is studied for the first time. The atomically-thin switching medium layer is formed through a unique process involving the oxidation of two-dimensional hexagonal boron nitride. It is observed that the confinement of the filament to atomic-scale thickness results in distinct atomic kinetics for the filament formation. The resulting memristor device can operate at sub-picoampere on-state current, less than 1 V operating voltage, and energy consumption per bit of less than 10 fJ for SET and less than 1 fJ for RESET. The device brings us one step closer towards realizing memristive electronic computation at the energy level of sub-femtojoule per bit. It is promising for applications in many emerging applications that desire ultra-low power operation, such as binary synapse in neuromorphic computation systems.

**Experimental Section**

*Few-layer hBN Flake Preparation and Oxidation:* Few-layer h-BN flakes were exfoliated onto the Si/SiO$_2$ substrate with 285 nm SiO$_2$ layer. After THE exfoliation and layer number identification, samples with bilayer and trilayer hBN were treated with low power remote oxygen plasma (Evactron® Plasma Decontaminator™ ) to fully oxidize the sample. An RF power of 12 W with 200 mTorr O$_2$ pressure and 20-30 seconds treatment time was applied. The samples were kept at least ~10 cm away from the plasma source to avoid damaging the surface morphology. Thicker h-BN flakes (>3 layers) were oxidized by Reactive Ion Etching (Oxford Instruments, Plasmalab 80 RIE system) under 200 mTorr O$_2$ pressure, 10-15 W RF power, and 15-30 s treatment time. The oxidized sample was then deterministically transferred onto a multilayer graphene flake, which serves as the inert bottom electrode of the device.

*AFM and Raman spectroscopy:* The AFM image was captured using a Bruker Dimension Icon System under an automatic peak force tapping mode. The Raman spectroscopy was measured with a Renishaw InVia spectrometer with a 532 nm incident laser. 50X objective and 100 uW laser power were used to avoid damaging the samples.

*STEM imaging and EELS measurements:* Cross-sectional STEM samples were prepared using an FEI Helios660 dual beam FIB/SEM system. In situ FIB lift-out method using FEI Helios 660 FIB/SEM was used to prepare the cross-sectional STEM samples with a protective C layer. Low energy (2 keV) Ga$^+$ ion beam was used for final milling. A Hitachi HD2700 and JEOL

ARM 200CF with an aberration corrector were used for high-angle annular dark-field (HAADF) scanning transmission electron microscopy (STEM) images were obtained with collection angles ranging from 68 to 280 mrad. EELS data were obtained with the JEOL ARM 200CF equipped with a cold field-emission gun. The energy dispersion and energy resolution were 0.25 eV/channel and better than 1 eV, respectively. Atomic ratios derived from the EELS data was performed by the quantification routine provided in Gatan Digital Micrograph[TM].

*Device Fabrication and Electrical Characterization:* The device was patterned by a standard EBL technique and the metal electrodes were deposited by electron beam metal evaporation. We used 40 nm Ag as the top electrode, with another 40 nm Au on top. Ag was deposited onto the BNO$_x$ layer at a slow rate (~ 0.5 A/s) to minimize metal ion penetration. All the devices were measured by a Keysight B1500A parameter analyzer using a lakeshore probe station at ambient environment.

*Kinetic Monte Carlo Simulations:* In two-dimensional (2D) KMC simulations, the hopping rate between the adjacent sites is described by a well-known thermodynamic relation,

$$\gamma_h = \gamma_0 exp\left(-\frac{E_a - \beta a_0 \varepsilon}{kT}\right) \quad (1)$$

where $E_a$ is the barrier height for a hopping mechanism, $a_0$ is the grid spacing, kT is the thermal energy, $\varepsilon$ is the electric field, $\gamma_0 = 10^{12} s^{-1}$ is the hoping rate constant in the order of $kT/h$ where $h$ is the Planck constant. $\beta$ is a unitless parameter that defines the efficiency at which the electric field assists in lowering the hopping/oxidation barrier height. A rectangular grid is used in the simulation by assuming the hopping sites are abundant for simplicity. The oxidation process at the top active electrode (AE) is described by the same thermodynamic relation in Eq. 1, in which $E_a$ is the oxidation barrier height and the electric field $\varepsilon$ is taken at the AE surface. The filament formation is assumed to be limited by the oxidation and hopping processes, rather than reduction at the bottom inactive electrode. The potential distribution is calculated from a resistive network,[36] in which the resistance value between two metallic adjacent sites is low, and it is high otherwise.

The concise model above has made a simplified assumption on the spatial distribution of hopping sites and has uncertainty of the hopping/ionization barrier height and rate constant parameters, but it is sufficient to study qualitative scaling trends as a function of the oxide thickness, which is insensitive to the exact values of the parameters.

**Acknowledgement**

This work was supported by Army Research Office (Grant no. W911NF-16-1-0435), Air Force Office of Scientific Research FATE MURI program (Grant no. FA9550-15-1-0514) and National Science Foundation (Grant no. CCF-1618038 and CCF-1618762). The work at Brookhaven National Laboratory was supported by the US DOE Basic Energy Sciences, Materials Sciences and Engineering Division, as well as the Center for Functional Nanomaterials which is U.S. DOE Office of Science Facilities, operated at Brookhaven National Laboratory under Contract No. DE-SC0012704. This research was supported in part by the Department of Energy (DOE) Award No. DE-FG02-07ER46376 (L.S).

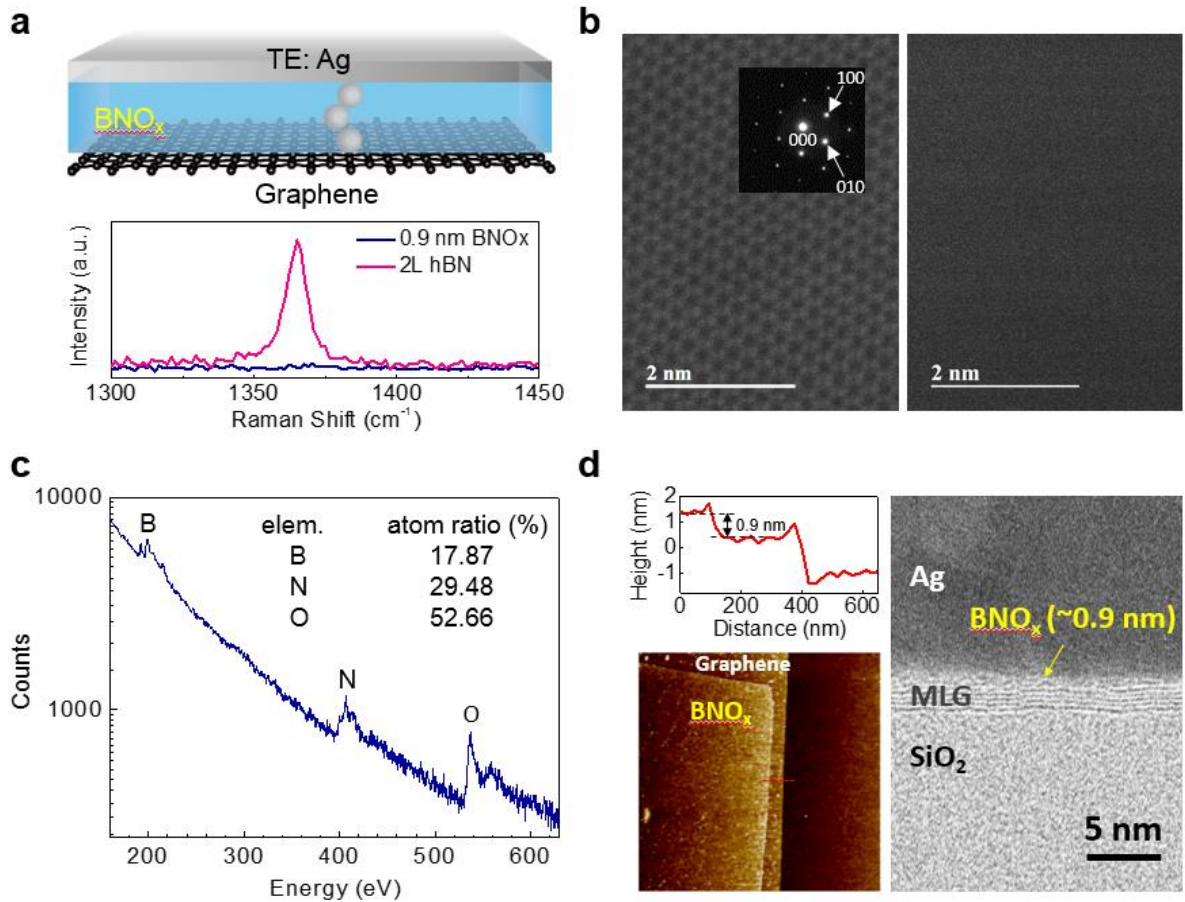

**Figure 1: Device structure and BNO$_x$ characterization.** (a) Upper panel: The schematic of an atomically-thin memristive device. A 0.9 nm thick BNO$_x$ is sandwiched between the multi-layer graphene bottom electrode and the Ag top electrode. Bottom panel: Raman spectra of a bi-layer h-BN flake, before (red) and after (blue) oxygen plasma treatment. (b) High-resolution STEM images show the crystalline lattice of h-BN before the oxygen plasma treatment (left panel) and the sample becomes amorphous after the treatment (right panel). The scale bar is 2 nm. The insets of left panel is the electron diffraction pattern of the crystalline h-BN, where crystal planes and inter-planar spacing are indicated by Miller indices. (c) Electron energy loss spectroscopy (EELS) spectrum of a BNO$_x$ flake. The element atom ratio confirmed the abundance of oxygen in the sample after the oxygen plasma treatment. (d) The lower left panel is the AFM image of a BNO$_x$ stacked on a 5-layer graphene, of which the height profile is shown at the upper left panel. The right panel is the cross-section STEM image of a device made from the sample in the left panel. The layered structure of 5-layer graphene and the amorphous morphology of $BNO_x$ can be clearly observed.

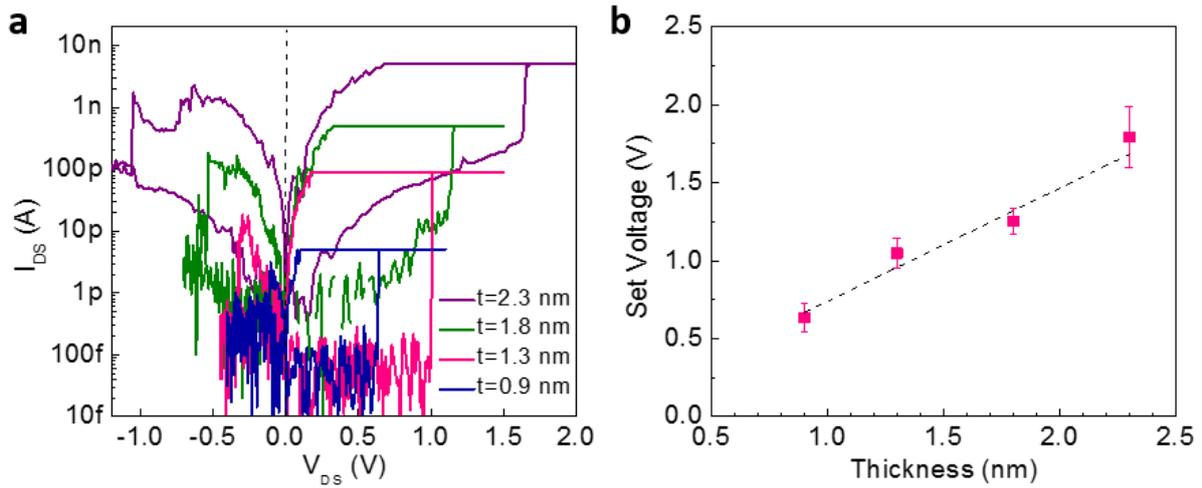

**Figure 2: Thickness dependent ultra-low power filamentary switching.** (a) The set-reset I-V characteristics of devices with 0.9 nm, 1.3 nm, 1.8 nm, and 2.3 nm $BNO_x$ layer thicknesses measured under current compliances of 5 pA, 90 pA, 500 pA, and 2 nA, respectively. (b) The linear dependence of the device set voltage on the $BNO_x$ layer thickness. For each device, several (>5) switching cycles were measured to obtain a statistical distribution, indicated by the error bar.

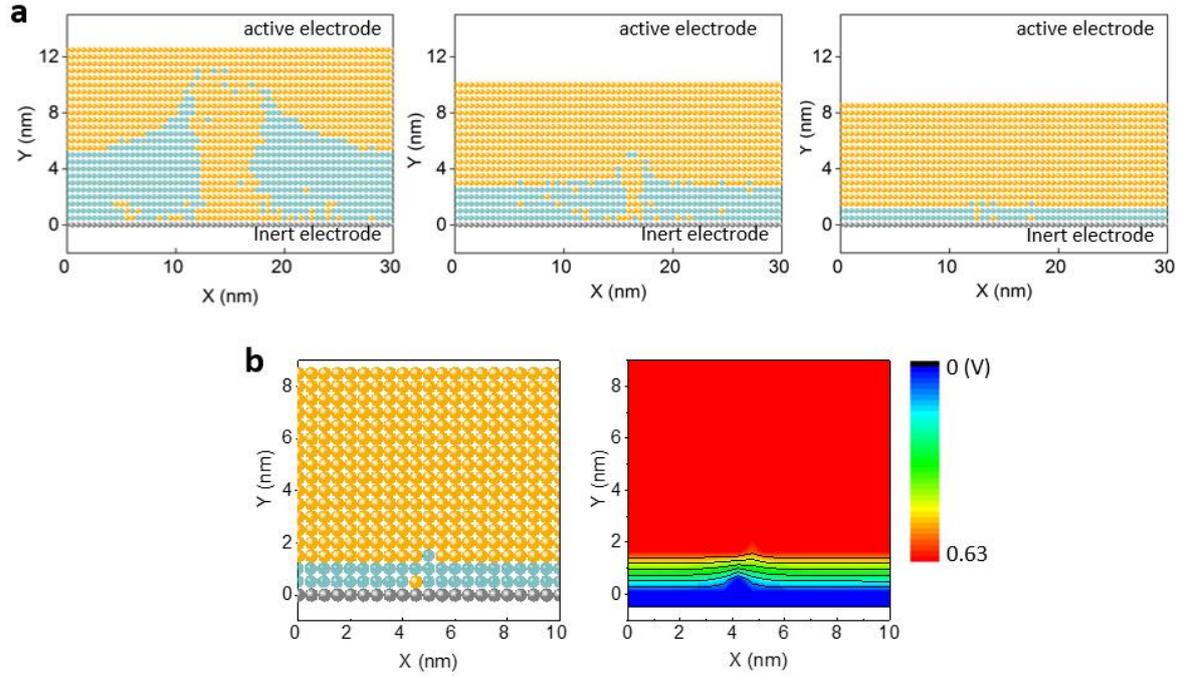

**Figure 3: Monte Carlo Simulation of the Filament Morphology.** (a) KMC simulations of the filament formation process in mediums with 4.5 nm, 2.25 nm, and 0.9 nm oxide sites. The filament in the 4.5 nm medium is wider because it grows both vertically and laterally, while the simulation on 0.9 nm switching medium shows the possibility of forming a single atomic chain filament. The applied bias for the 4.5 nm and 0.9 nm structure is 3.15 V and 0.63 V, respectively. (b) Electrical potential profile during the filament growth in the 0.9 nm switching medium, as shown in the left panel. The applied voltage is 0.63V. The right panel demonstrates the local electrical field enhancement induced by a hoping of Ag cation, which is a situation depicted in the left panel.

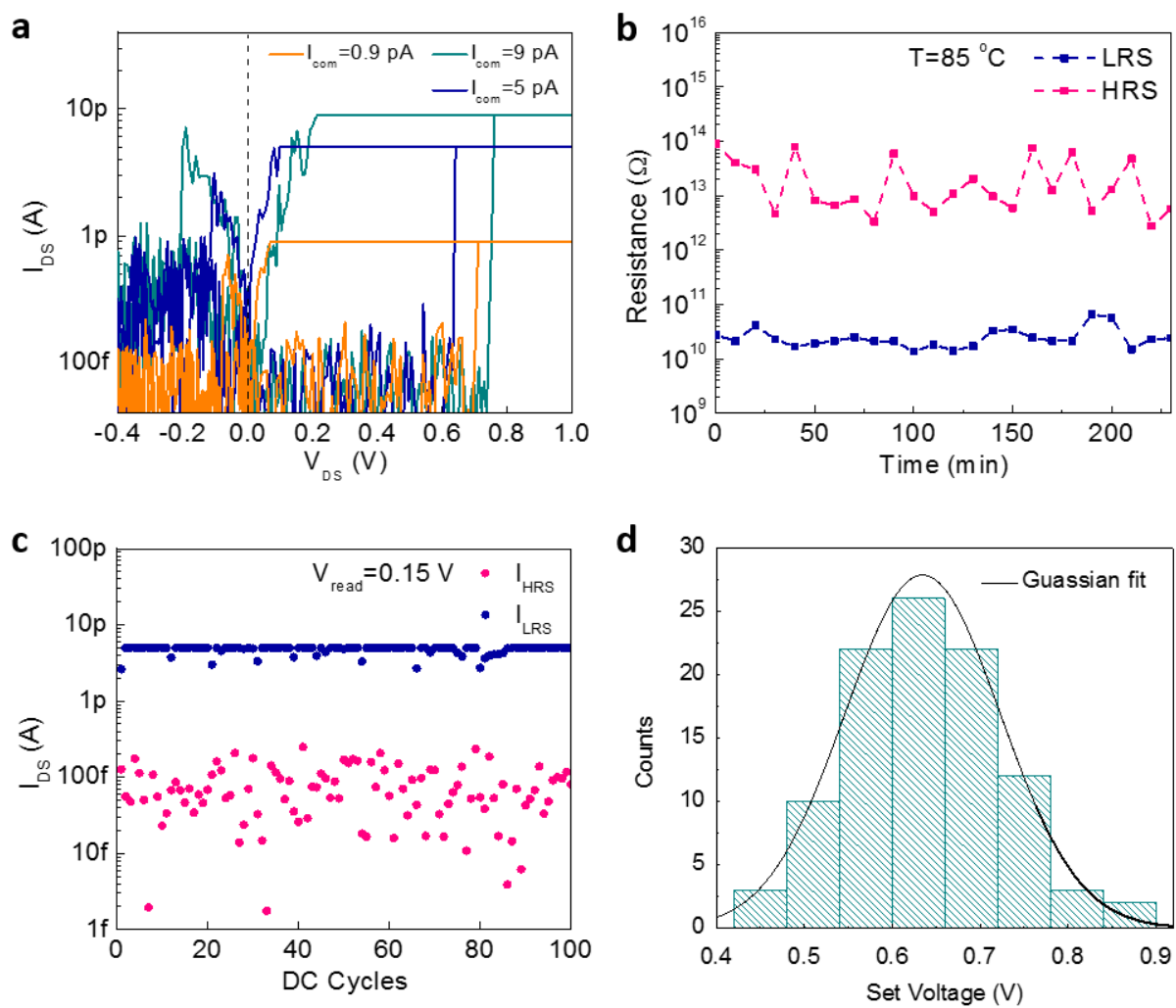

**Figure 4: Characteristics of a device with 0.9 nm BNO$_x$.** (a) Set-reset I-V characteristics of a device with 0.9 nm BNO$_x$ switching layer subject to different current compliances of 0.9 pA, 5 pA, and 9 pA. (b) Data retention measured at 85 °C. Both HRS and LRS resistances were measured with a 0.18 V read voltage. (c) The current read at 0.15 V over 100 continuous switching cycles. (d) The statistical distribution of the set voltage for 100 switching cycles and the Guassian fit curve. The set voltages exhibit a normal distribution centered at 0.63 V, with a standard deviation of 0.088 V.